\documentclass{mn2e}
\usepackage{times}
\usepackage{graphicx}

\newbox\grsign \setbox\grsign=\hbox{$>$} \newdimen\grdimen
\grdimen=\ht\grsign
\newbox\simlessbox \newbox\simgreatbox \newbox\simpropbox
\setbox\simgreatbox=\hbox{\raise.5ex\hbox{$>$}\llap
     {\lower.5ex\hbox{$\sim$}}}\ht1=\grdimen\dp1=0pt
\setbox\simlessbox=\hbox{\raise.5ex\hbox{$<$}\llap
     {\lower.5ex\hbox{$\sim$}}}\ht2=\grdimen\dp2=0pt
\setbox\simpropbox=\hbox{\raise.5ex\hbox{$\propto$}\llap
     {\lower.5ex\hbox{$\sim$}}}\ht2=\grdimen\dp2=0pt
\def\simgreat{\mathrel{\copy\simgreatbox}}
\def\simless{\mathrel{\copy\simlessbox}}

\def\NH1{{$N_{\rm HI}~$}}

\def\Rsun{\hbox{$\rm\thinspace R_{\odot}$}}
\def\Msun{\hbox{$\rm\thinspace M_{\odot}$}}

\title[Near-IR spectroscopy of PG\,1234+482]{Near-infrared spectroscopy of the very low mass companion to the hot DA white dwarf PG\,1234+482}

\author[P. R. Steele et al.]{P. R. Steele$^{1}$\thanks{E-mail:
prs15@star.le.ac.uk} M. R. Burleigh$^{1}$ P. D. Dobbie$^{2}$ and M. A. Barstow$^{1}$ \\
$^{1}$Department of Physics and Astronomy, University of Leicester, University Road, Leicester LE1 7RH, UK\\
$^{2}$Anglo-Australian Observatory, PO Box 296, Epping, Sydney, NSW 1710, Australia}
\begin{document}

\date{March 2007}

\pagerange{\pageref{firstpage}--\pageref{lastpage}} \pubyear{2007}

\maketitle

\label{firstpage}

\begin{abstract}
We present a near-infrared spectrum of the hot ($T_{\rm eff}$\,$\approx$\,55,000\,K) DA white dwarf PG\,1234+482. We confirm that a very low mass companion is responsible for the previously recognised infrared photometric excess. We compare spectra of M and L dwarfs, combined with an appropriate white dwarf model, to the data to constrain the spectral type of the secondary. We find that uncertainties in the 2MASS $HK$ photometry of the white dwarf prevent us from distinguishing whether the secondary is stellar or substellar, and assign a spectral type of L0$\pm$1 (M9-L1).Therefore, this is the hottest and youngest ($\approx 10^6$~yr) DA white dwarf with a possible brown dwarf companion.
\end{abstract}


\section{Introduction}
Observations of substellar companions to white dwarfs allow the investigation of a variety of aspects of binary formation and evolution. Since a white dwarf is up to $\sim$\,10,000~times fainter than its progenitor, the contrast gain also facilitates direct detection of very low mass secondaries. White dwarfs with brown dwarf secondaries can be used to place constraints on the fraction of their main sequence progenitors with substellar companions. For example, radial velocity and imaging surveys indicate a discrepancy between the brown dwarf companion fraction at small separations ($1\pm1$\% at $<10$\,AU, Marcy and Butler 2000; McCarthy \& Zuckerman 2004) and large radii ($a>1000$\,AU; $10-30$\%; Gizis et al. 2001). The closest white dwarf $+$ brown dwarf binaries might also represent either another channel for cataclysmic variable (CV) evolution (Politano 2004) or the end state of CV evolution, in which the secondary has become highly evolved through mass transfer. In close detached binaries, the brown dwarf is expected to be irradiated by the white dwarf's high UV flux, possibly leading to substantial differences between the ``day'' and ``night'' hemispheres. Such contrasts have recently been reported in several hot Jupiters (e.g. HD\,189733b, Knutson et al. 2007). Finally, we note that there are few observational constraints on brown dwarf evolutionary models at large ages, such as might be expected for most white dwarfs ($>1$~Gyr). Thus wide, detached white dwarf $+$ brown dwarf binaries may provide ``benchmarks'' for testing these models (Pinfield et al. 2006).
\\ \indent However, brown dwarf companions to white dwarfs are rare (Farihi et al. 2005). To date only three such systems have been confirmed; GD165 (DA+L4, Becklin \& Zuckerman 1988), GD1400 (DA$+$L6/7, Farihi \& Christopher 2004; Dobbie et al. 2005) and WD0137-349 (DA$+$L8, Burleigh et al. 2006; Maxted et al. 2006). GD165 is a widely separated system (120AU) whereas WD0137-349 is a close binary (P$_{\rm orb}$\,$=$\,116 minutes). The separation of the components in the GD1400 system are currently unknown. At $T_{\rm eff}$\,$\approx$\,55,000\,K, PG\,1234$+$482 is significantly hotter than GD165 ($T_{\rm eff}$\,$\approx$\,12,000\,K, McCook \& Sion 1987), GD1400 ($T_{\rm eff}$\,$\approx$\,11,580\,K, Farihi \& Christopher 2004), and WD0137-349 ($T_{\rm eff}$\,$\approx$\,16,500\,K, Maxted et al. 2006). Although there are many candidate substellar mass secondaries in CV's, it was only recently that one was finally confirmed (SDSS\,103533.03+055158.4, Littlefair et al. 2006), though not through a direct spectroscopic detection.
\begin{figure*}
\includegraphics[angle=270,width=170mm]{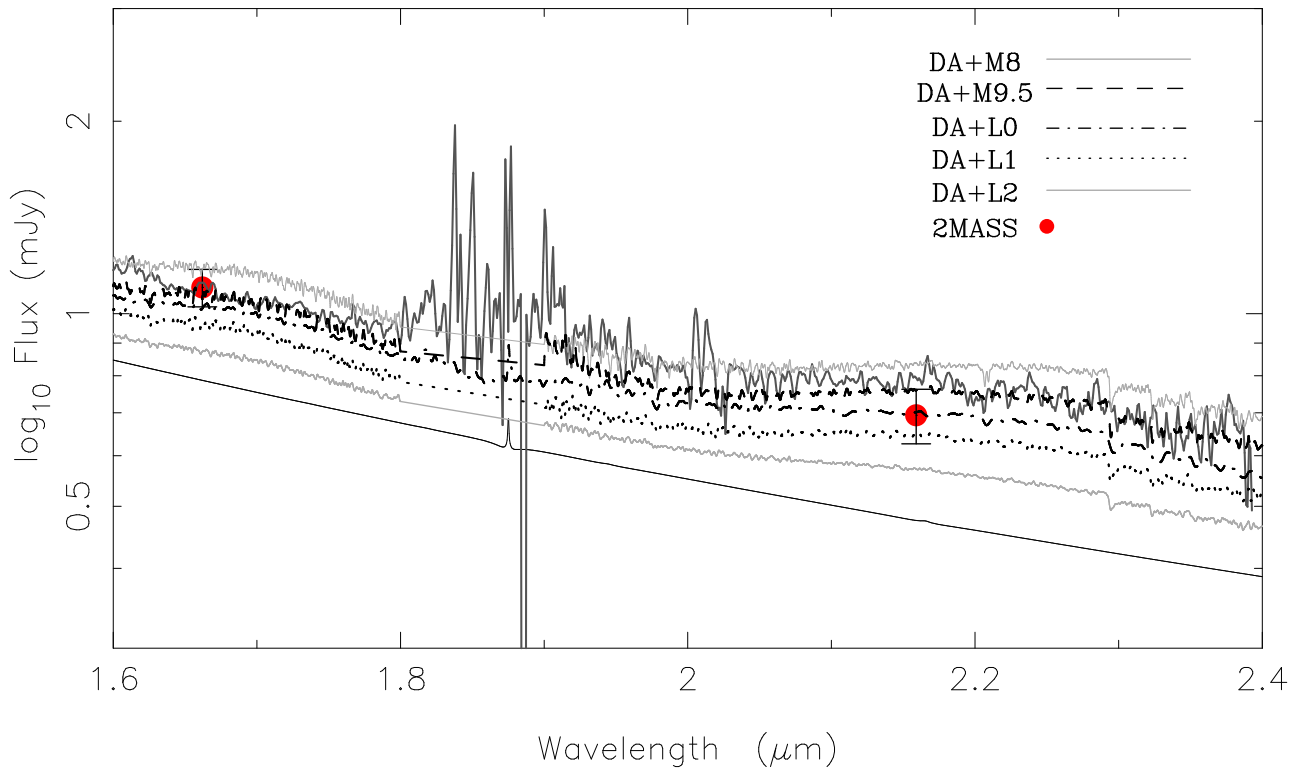}
\caption{Observed near-infrared spectrum of PG\,1234+482 scaled to the 2MASS $H$ flux. The predicted white dwarf model alone is shown by the solid line. We also compare the data to the white dwarf model combined with late M and early L dwarfs, all scaled to a distance of 144\,pc (Liebert, Bergeron \& Holberg 2005). The upper and lower grey spectra are PG\,1234$+$M8 and PG\,1234$+$L2 spectra respectively. The 1.87$\mu$m H Paschen $\alpha$ emission line in the predicted white dwarf spectrum is due to non-LTE effects in the upper atmosphere.}
\end{figure*}
\\ \indent Clues to the existence of other white dwarf $+$ brown dwarf binaries might be provided by the degenerate stars themselves. White dwarfs are expected to have either pure-H or pure-He atmospheres. However, in some apparently isolated white dwarfs there are unusually high metal abundances. This is somewhat unexpected as the gravitational settling times of such elements are much less than the cooling timescale of the white dwarf. For example, in H-rich DAZ white dwarfs with $T_{\rm eff}\simless$\,20,000\,K Ca and Mg are seen in optical spectra, despite these elements having a residence time in the atmosphere of days. This strongly implies that the polluting material is being directly accreted from some unseen source. Zuckerman et al. (2003), Gianninas, Dufour \& Bergeron (2004), and Kilic \& Redfield (2007) argue that interstellar material is an unlikely source of accretion, implying that the metals are being accreted from a source associated with the white dwarf itself. The detection of dust and gas disks around some DAZs provides the source for this material in these cases (e.g. GD362, Becklin et al. 2005; Kilic et al. 2005; SDSSJ\,104341.53+085558.2, Gaensicke, Marsh \& Southworth 2007), but Zuckerman et al. (2003) have also noted the high frequency of DAZ stars which are part of known binary systems and suggest that wind-driven mass loss from the companion may be responsible for a proportion of the observed DAZ stars. 
\\ \indent Barstow et al. (2003) used far-UV spectra to measure the metal abundances for a group of hotter DA white dwarfs in the range 20,000\,K$< T_{\rm eff}<$\,110,000\,K. They showed that the photospheres of all DAs $\simgreat$\,50,000\,K contain heavy elements which can be adequately supported against gravitational settling by radiation pressure. In most cases these metals presumably remain from the pre-white dwarf evolution of the star. Therefore, a difficulty may arise in easily identifying candidate hot stars that might be accreting small amounts of material from an unseen dust or gas disk, or the wind of a close companion, unless the latter is of spectral type mid-M or earlier. For example, the hot DA white dwarfs Feige~24\, ($T_{\rm eff}$\,$\approx$\,60,000\,K, Vennes et al. 2000; Barstow et al. 2003) and RE\,J\,0720$-$318 ($T_{\rm eff}$\,$\approx$\,55,000\,K Dobbie et al. 1999; Vennes, Thorstensen \& Polomski 1999) have close, early-M companions which are possibly supplying at least a fraction of the heavy elements detected in the white dwarfs' photospheres. It is plausible that a proportion of the apparently single metal-polluted hot DA white dwarfs may also be accreting from as yet unseen low mass companions (Dobbie et al. 2005).
\\ \indent PG\,$1234+482$ (hereafter PG\,1234) is a hot ($T_{\rm eff}$\,$\approx$\,55,000\,K) DA white dwarf with some evidence for metal pollutants in its photosphere. Jordan, Koester \& Finley (1996) reported the detection of Fe in an {\it Extreme Ultraviolet Explorer} spectrum of the star, and Wolff et al. (1998) give the overall metallicity as 20\% that of the archetypal metal-rich hot DA G191$-$B2B using the same data (a pure-H atmosphere over-predicts the EUV continuum flux and metal opacities are required to satisfactorily model these data). But Barstow et al. (2003) did not detect any heavy elements in a noisy {\it International Ultraviolet Explorer} far-UV spectrum, and only give upper limits on the abundances of C, N, O, Si, Fe and Ni. A quick analysis of the {\it FUSE} spectrum of PG\,1234 also failed to reveal any obvious photospheric metal lines.
\\ \indent PG\,1234 was first observed in the near-IR by Green, Ali \& Napiwotzki (2000), who reported a small $1.3\sigma$ $K$-band excess and thus did not claim any evidence for a companion. Debes, Sigurdsson \& Woodgate (2005) later noted significant $H$ and $K$-band excess from the more precise 2MASS photometry and  suggested the presence of a companion spectral type M8V. More recently Mullally et al. (2006) measured a mid-infrared excess in two Spitzer IRAC bands ($4.5\mu$m and $8.0\mu$m). They modelled the excess using a companion with $T_{\rm eff}<2000$K, assigning a spectral type of early L.\\ \indent Here we report near-IR spectroscopy of PG\,1234 to confirm the presence of the very low mass companion and to better constrain its spectral type.

\section{Observations and Data Reduction} 
We observed PG\,1234 on March 5th 2007 using the Long-slit Intermediate Resolution Infrared Spectrograph (LIRIS) during service time on the 4.2m  William Hershel Telescope (WHT). LIRIS is a near-IR imager/spectrograph which uses a $1024 \times 1024$ pixel array  HAWAII detector. The pixel scale is 0.25$\arcsec$/pixel, giving a field view of $4.27^{\prime} \times 4.27^{\prime}$. Data were acquired using the $HK$ grism providing a wavelength coverage of $1.39-2.42\mu$m. Observations were taken  using the standard technique of nodding the point source targets along the spectrograph slit in an ABBA pattern. For PG\,1234 $18 \times 100$s exposures were taken (9 AB pairs) for a total exposure time of 1800s, followed by $4 \times 10$s exposures of an A3V telluric standard. The average airmass over the course of the observations was 1.06.  
\begin{figure*}
\includegraphics[angle=270,width=120mm]{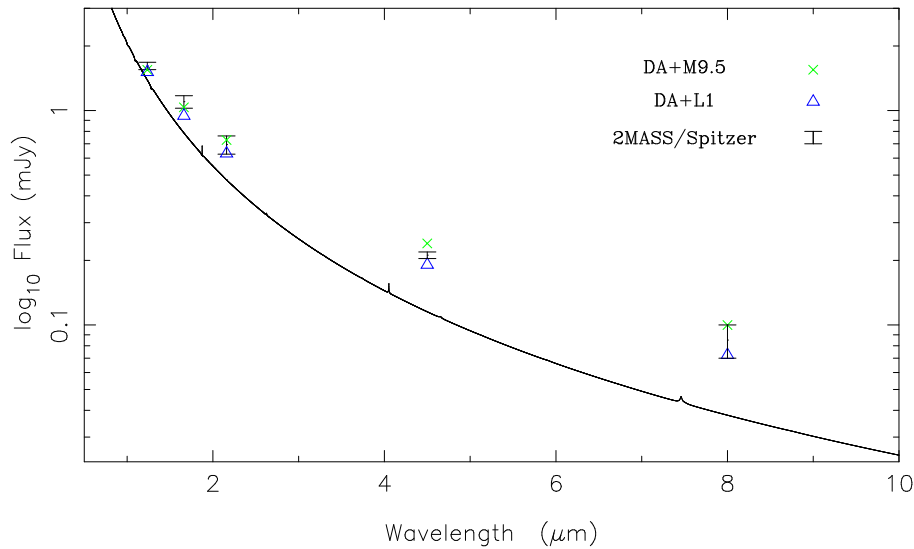}
\caption{2MASS and Spitzer IRAC (4.5$\mu$m and 8.0$\mu$m) photometric fluxes of PG\,1234+482 with M9.5 and L1 fluxes scaled appropriately. The symbols for the combined WD $+$ M and L dwarf models do not represent the actual errors which can be seen in table 1.}
\end{figure*}
\begin{table*}
\caption{Photometric fluxes in mJy of combined PG\,1234 (Predicted) $+$ M and L dwarfs compared to those of PG\,1234 (Observed).}
\begin{tabular}{llllcccrr}
\hline
& & \multicolumn{3}{|c|}{2MASS} & \multicolumn{2}{|c|}{Spitzer IRAC Channel}\\
Name & Spectral Type & $J$(error) & $H$(error) & $K_{\rm s}$(error) & 4.5$\mu$m(error) & 8.0$\mu$m(error) \\
\hline
PG\,1234 (Observed)    & DA   & 1.628(28) & 1.067(21) & 0.707(13) & 0.2116(75) & 0.085(15) \\
PG\,1234 (Predicted) $+$ BRI 0021-0214 & DA $+$ M9.5 & 1.549(27) & 1.035(20) & 0.729(13) & 0.240(10)   & 0.010(18) \\
PG\,1234 (Predicted) $+$ 2MA 1439+1929 & DA $+$ L1.0 & 1.512(26) & 0.946(19) & 0.631(12) & 0.191(8)   & 0.072(12) \\
\hline
\end{tabular}
\end{table*}
To reduce the data we first corrected the bottom-left quadrant pixel 'scrambling' (the image is constructed after readout with a 1-pixel dislocation) in all images using the \begin{it} lcpixmap \end{it} function of the IRAF package LIRISDR \footnote{http://www.iac.es/proyect/LIRIS/index.html}. We then applied standard reduction techniques using software routines in the STARLINK packages KAPPA and FIGARO. In brief, a bad pixel map was constructed and applied to all the data. The science, standard star and arc lamp spectral images were flat fielded with a normalized response map. Difference pairs were then assembled from the science and standard star images and any significant remaining sky background was removed by subtracting linear functions, fitted in the spatial direction, from the data. The spectra of the white dwarf and the standard star were then extracted and assigned the wavelength solution derived from the arc spectrum. Any intrinsic features of the standard star's energy distribution were identified by reference to the near-IR spectral atlas of fundamental MK standards (Wallace et al. 2000, Meyer et al. 1998, Wallace \& Hinkle 1997) and were removed by linearly interpolating over them. The spectrum of the white dwarf was then co-aligned with the spectrum of the standard star by cross-correlating the telluric features present in the data. The science spectrum was divided by the standard star spectrum and multiplied by a blackbody with $T_{\rm eff}$ of the standard. Finally, the flux levels were scaled to achieve the best possible agreement between the spectral data and the $H$ and $K_{s}$ photometric fluxes of the object derived from the 2MASS All Sky Data Release Point Source Catalogue magnitudes (Skrutskie et al. 1997)

\section{Analysis}
To assist in spectrally typing the companion, we compare the data to combined white dwarf and M/L dwarf models.

We have generated a pure-H synthetic spectrum for PG\,1234 at T$_{\rm eff}$\,$=$\,55,040\,K and log$g$\,$=$7.78 (Liebert, Bergeron \& Holberg 2005) using the plane-parallel, hydrostatic, non-local thermodynamic equilibrium (non-LTE) atmosphere and spectral synthesis codes TLUSTY (v202; Hubeny 1988; Hubeny \& Lanz 1995) and SYNSPEC (v49). The synthetic flux as been normalized to $V=14.45$ (Liebert, Bergeron \& Holberg 2005).

\begin{table}
\caption{The lowest mass detached companions to white dwarfs.}
\begin{tabular}{llllcccrr}
\hline
Name & SpTypes & Separation &Ref.\\
\hline
WD\,2151-015 & DA$+$M8 & 23AU & 1 \\
WD\,2351-335 & DA$+$M8 & 2054AU & 2 \\
WD\,1241-010 & DA$+$M9 & 284AU & 2 \\
PG\,1234+482 & DA$+$L0$\pm$1 & Unresolved & 3 \\
GD\,165 & DA$+$L4 & 120AU  &  4 \\
GD\,1400 & DA$+$L6/7 & Unresolved & 5,\,6 \\
WD\,0137-349 & DA$+$L8 (0.053\Msun) & 0.65\Rsun (P=116 mins) & 7,\,8 \\
\hline
\end{tabular}
\\
1.\,Farihi, Hoard \& Wachter (2006) 2.\,Farihi, Becklin \& Zuckerman (2005)
3.\,This work 4.\,Becklin \& Zuckerman (1988) 5.\,Farihi \& Christopher (2004)
6.\,Dobbie et al. (2005) 7.\,Maxted et al. (2006) 8.\,Burleigh et al. (2006) 
\end{table}
\begin{table}
\caption{The lowest mass companions to CV's}
\begin{tabular}{llllcccrr}
\hline
Name & SpTypes & Period & Ref. & & &\\
\hline
OY Car & CV$+$M6 & 91\,mins & 1 &&&\\
EX Eri & CV$+$L0.084\Msun\,Star & $\sim$90\,mins & 2 &&&\\
SDSS\,1212 & DA$+$L8-T2 & 88.4\,mins & 3,\,4 &&&\\
SDSS\,1035 & CV$+$0.052\Msun\, BD & 82\,mins & 5 &&&\\
SDSS\,1507 & CV$+$0.056\Msun\, BD & 66.61\,mins & 6 &&&\\
\hline
\end{tabular}
\\
1.\,Wood \& Horne (1990) 2.\,Feline et al. (2004) 3.\,Burleigh et al. (2006) 
4.\,Farihi, Burleigh \& Hoard (2007) 5.\,Littlefair et al. (2006) 
6.\,Littlefair et al. (2007)
\end{table}

Empirical companion models have been constructed using the near-IR spectra of M and L dwarfs from the IRTF spectral library (Cushing et al. 2005; Rayner et al. in prep.). The fluxes of the empirical models have been scaled to a level appropriate to a location at $d=10$\,pc using distances estimated from the parallax of each object. Subsequently, these fluxes have been re-scaled to be consistent with the Liebert, Bergeron \& Holberg (2005) distance estimate of $144$\,pc. The final distance scaled models were then added to our synthetic white dwarf spectrum.

We also compare the PG\,1234 Spitzer photometry of Mullally et al. (2006) to archival Spitzer IRAC photometry of observed M and L dwarfs (Table 1, Patten et al. 2006). Again, these values have been appropriately scaled to the distance given previously. We added these fluxes to expected values for PG\,1234 calculated by integrating our synthetic spectrum folded though the appropriate filter transmission response. 

\section{Results}
Figure 1 shows our extracted spectrum for PG\,1234. It is not possible to scale the spectrum to match the $H$ and $K$ 2MASS photometry simultaneously, although when scaling to one the other is only off by $<2\sigma$. This may indicate either a residual error in the reduction of the spectrum or errors in the 2MASS photometry. At $K_s = 14.937 \pm 0.106$ PG\,1234 does not meet the 2MASS Point Source Catalog level 1 requirements (S/N $> 10$) and Tremblay \& Bergeron (2007) have shown that lower quality 2MASS data should be treated with caution when interpreting near-infrared excesses to white dwarfs. In Figure 1 we show the data scaled to the $H$ photometry. Figure 1 also shows combined white dwarf $+$ dwarf spectra of spectral types M9.5, L0 and L1. The light grey spectra above and below are an M8 and L2 respectively added to show our data do not match either of these types. When normalized to the $H$-band flux the observed spectrum is best approximated by a WD$+$M9.5 companion, whereas if normalized to the $K$-band flux the spectrum would be closer to that of a WD$+$L0 or WD$+$L1.
\\ \indent Figure 2 shows the 2MASS $JHK$ and the Spitzer IRAC 4.5$\mu$m and 8.0$\mu$m photometry (Mullally et al. 2006) of PG\,1234 with the combined predicted PG1234 fluxes $+$ BRI 0021-0214 (M9.5) and 2MA 1439+1929 (L1). It can be seen that the 2MASS fluxes are more closely approximated by the WD$+$M9.5 spectral type, whereas Spitzer fluxes are overpredicted by our WD$+$M9.5 model, but are underpredicted by our WD$+$L1 model. Thus the Spitzer data are likely best matched with a WD$+$L0.

\section{Conclusions}
We have attempted to determine the spectral type of the low mass companion to PG\,1234+482 using $H$ and $K$-band spectroscopy obtained from the WHT LIRIS instrument and the Spitzer IRAC $4.5\mu$m and $8.0\mu$m photometry. Due to uncertainty in the 2MASS photometry used to place these data on an absolute flux scale, we estimate the spectral type as L0$\pm$1 (M9-L1), making PG\,1234 the hottest and youngest ($\rm t_{cool}$$\approx 10^6$~yr; Liebert, Bergeron \& Holberg 2005) DA white dwarf with a possible brown dwarf companion. 

Whether the companion is substellar then depends on its age, which we can estimate as follows. The mass of the progenitor to PG\,1234 can be estimated using the initial-final mass relationship of Dobbie et al. (2006), which holds down to an initial mass of 2.7\Msun, but has recently been shown to extend down to an initial mass of 1.6\Msun\,by Kalirai et al. (2007). The mass of PG\,1234 is 0.61$\pm$0.02\Msun \ (Liebert, Bergeron \& Holberg 2005) yielding an initial mass of 2.4$\pm$0.1\,\Msun. An approximate main sequence lifetime can then be calculated from:$$\rm t_{\rm ms}=10\,[(M_{*}/\rm\thinspace M_{\odot})^{-2.5}]\,\rm Gyr \, (\rm Wood\, 1992)$$ Therefore, the age of the system is $\approx1$~Gyr as the cooling age is negligible.  

We can estimate the effective temperature of an object on the substellar boundary using the DUSTY models of Chabrier et al. (2000) and Baraffe et al. (2001). At an age of 1\,Gyr and the commonly used upper mass limit for brown dwarfs of 0.075\Msun, we expect an effective temperature of $\approx 2200$\,K. At this temperature the observations of Vrba et al. (2004) suggest a spectral type of L1-L2, and Golimowski et al. (2004) an L1. Using the empirical formula of Stephens et al. (2001):$$\rm T_{\rm eff}=2220-100\rm L_{\rm n}\,\,\rm where\,\, L_{\rm n}=\rm L0-L8$$we would expect a spectral type of L0-L1. Thus, the expected spectral type of a 1\,Gyr object on the substellar boundary is $\approx$\,L1$\pm$1. 

Therefore, we do not have sufficient evidence to state conclusively if the companion is stellar or substellar in nature - i.e. a brown dwarf. To further constrain the spectral type we suggest higher resolution, higher S/N $H$ and more importantly $K$-band spectroscopy. An accurate spectral type could then be determined from the relative strength of the CO absorption edges at 2.3$\mu$m. 

The confirmation of a very low mass, possibly substellar, companion to PG\,1234 leads us to speculate whether the metals in the hot white dwarf's atmosphere are at least in part being accreted from the companion's wind. To answer this we will need to constrain the separation of the pair, by either high resolution imaging and/or radial velocity measurements. Farihi et al. (2005) did not resolve the two, indicating that the separation is less than or approximately 1$\arcsec$. Consideration of the post-main sequence evolution of these systems and observations by Farihi, Wachter \& Hoard (2006) suggests that there may be a bimodal distribution of orbital separations among binaries containing at least one white dwarf: wide pairs with orbits $>10$\,AU and very close systems ($<$~few solar radii) in which the companion was dragged in during the common envelope phase. The failure to resolve the system by Farihi et al. (2005) might be indicating that PG\,1234 is a close binary, but we caution that the pair also may be aligned by chance.  Whether it is wide or close, and a possible survivor of common envelope evolution, PG\,1234$+$482\,B is one of the lowest mass companions to a white dwarf currently known. We list the known lowest mass detached companions to white dwarfs in Table 2, and the lowest mass companions to CV's in table 3.
 
\section*{Acknowledgements}
PRS is sponsored by STFC in the form of a studentship. MRB acknowledges the support of STFC in the form of an Advanced Fellowship. This publication makes use of data 
products from the Two Micron All Sky Survey, which is a joint project of the University of Massachusetts
and the Infrared Processing and Analysis Center/California Institute of Technology, funded by the National
Aeronautics and Space Administration and the National Science Foundation.

\appendix

\bsp

\label{lastpage}

\end{document}